# Acceleration of Linear Finite-Difference Poisson-Boltzmann Methods on Graphics Processing Units

Ruxi Qi, Wesley M. Botello-Smith, and Ray Luo*

Department of Molecular Biology and Biochemistry
University of California, Irvine, CA 92697-3900

* Please send correspondence to R. Luo. email: ray.luo@uci.edu; fax: (949) 824-9551.

Electrostatic interactions play crucial roles in biophysical processes such as protein folding and molecular recognition. Poisson-Boltzmann equation (PBE)-based models have emerged as widely used in modeling these important processes. Though great efforts have been put into developing efficient PBE 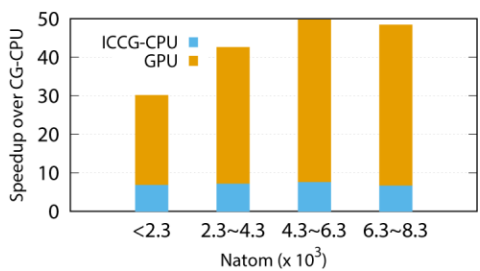 numerical models, challenges still remain due to the high dimensionality of typical biomolecular systems. In this study, we implemented and analyzed commonly used linear PBE solvers for the ever-improving graphics processing units (GPU) for biomolecular simulations, including both standard and preconditioned conjugate gradient (CG) solvers with several alternative preconditioners. Our implementation utilizes standard Nvidia® CUDA libraries cuSPARSE, cuBLAS, and CUSP. Extensive tests show that good numerical accuracy can be achieved given that the single precision is often used for numerical applications on GPU platforms. The optimal GPU performance was observed with the Jacobi-preconditioned CG solver, with a significant speedup over standard CG solver on CPU in our diversified test cases. Our analysis further shows that different matrix storage formats also considerably affect the efficiency of different linear PBE solvers on GPU, with the diagonal format best suited for our standard finite-difference linear systems. Further efficiency may be possible with matrix-free operations and integrated grid stencil setup specifically tailored for the banded matrices in PBE-specific linear systems.

# 1. Introduction

In recent years Poisson-Boltzmann equation (PBE)-based electrostatics modeling has gained wide acceptance in biomolecular applications, given the crucial roles played by the electrostatic interactions in biophysical processes such as protein-protein and protein-ligand interactions.[1] Due to the high dimensionalities of typical biomolecular systems, it is extremely important to increase the accuracy and efficiency of PBE models.[2]

For biomolecular applications, the PBE is impossible to be solved analytically, so that only numerical solutions are possible. Traditional numerical schemes include the finite difference method (FDM)[3] where difference grids are used to discretize the space and build up a set of linear/nonlinear equations from the PBE, and the finite-element method[4] where arbitrarily shaped biomolecules are discretized by using elements with a set of associated basis functions. The boundary element method is another alternative approach, in which only the surfaces of the molecules are discretized.[5] Numerical PBE methods have been applied to the prediction of pKa values for ionizable groups in biomolecules,[6] solvation free energies,[7] binding free energies,[8] and protein folding and design.[9]

Among these approaches, the FDM is most widely adopted and has been incorporated in programs such as DelPhi,[3a, 3c, 3j] UHBD,[3b, 3d] APBS,[3e, 3g] CHARMM/PBEQ,[3c, 3i] and Amber/PBSA.[2h, 3l-n, 10] The resulting algebraic systems are often solved by using conjugate gradient methods with or without preconditioners.[3b, 3k, 11] As computational studies shift to larger and more complex biomolecular systems, both the data storage and convergence rate become more challenging to address on traditional CPU platforms. These challenges are more pronounced when incorporating the PBE in typical molecular simulations involving thousands to millions of snapshots.

Recently, graphics processing units (GPU) have been used in a wide range of computational chemistry problems, including MD simulations[12] and ab initio quantum mechanical (QM) calculations[13] with impressive speedup. Different from CPUs that are designed for sequential execution, GPUs have a parallel architecture that is suited for high-performance computation with dense data parallelism, and have enjoyed rapid adoption over the last decade. A number of publications have also shown the use of GPUs to accelerate PBE linear systems for biomolecular systems and reported impressive speedup.[14] However, different from MD or QM simulations, various PBE solvers perform with markedly different efficiency.[3l, 3m] Simpler algorithms may be straightforward to be ported onto GPU platforms, but they may not be robust or efficient enough to begin with (i.e. they may be very slow to converge or need very high number of floating operation counts to achieve a given convergence criterion), particularly on very complex or large biomolecular systems. Therefore, a thorough analysis of existing algorithms on GPUs is a necessary step to realize markedly improved overall efficiency in numerical PBE solutions for biomolecular applications.

To date only the relatively simple successive over-relaxation (SOR) method was implemented on GPUs.[14] However, our prior algorithm analysis of SOR and other algorithms have shown its convergence rate is not among the best on CPU for large systems or tight convergence criterion even if it is a simple algorithm to implement.[3l, 3m] Furthermore, there are two additional disadvantages when porting the SOR method to GPUs. Firstly, a parallel SOR, such as red-black SOR, has to be used to utilize the parallel GPUs. However the red-black SOR has worse convergence rate than the original SOR due to its altered updating approach. Secondly, for most consumer-grade GPU cards, single precision operations are widely supported with high efficiency. Double precision operations are possible, but are at a significant performance disadvantage. Unfortunately use of single precision further deteriorates the convergence of red-black SOR whether it is on GPUs or on CPUs as our in-house testing has shown.

In this paper, we present the implementation and systemic assessment of four types of linear PBE solvers on GPUs using the Nvidia CUDA (Version 7.5) libraries. In the following the underlining linear systems solvers are first reviewed. This is followed by an assessment of the accuracy and efficiency observed for different implementations. The impact of matrix storage formats upon the computation efficiency is then discussed. Finally the memory usage on the GPUs is briefly addressed.

## 2. Methods

**2.1 Poisson-Boltzmann Equation**

In implicit solvent models, the solvent is treated as high dielectric continuum and the solute is approximated as low dielectric continuum with charges embedded inside. The PBE is then introduced to describe the electrostatic interactions in the heterogeneous dielectric environment, with the Boltzmann term describing the salt effect of a dissolved electrolyte. This gives the well-known non-linear PBE

$$\nabla \cdot \varepsilon(\mathbf{r})\nabla \phi(\mathbf{r}) + \lambda(\mathbf{r})\sum_i n_i q_i \exp[-q_i \phi(\mathbf{r})/kT] = -\rho(\mathbf{r}), \qquad (1)$$

where $\rho$ is the charge density, $\phi$ is the electrostatic potential, $\varepsilon$ is the dielectric constant, and $\lambda$ is a masking function for the Stern layer. All variables are functions of the spatial vector **r**. In the salt related term, $n_i$ is the number density of ion of type $i$ in the bulk solution, $q_i$ is the charge of the ion of type $i$, $k$ is the Boltzmann constant and $T$ is the temperature. When the term $q_i\phi(\mathbf{r})/kT$ is small, the PBE can be linearized into

$$\nabla \cdot \varepsilon(\mathbf{r})\nabla \phi(\mathbf{r}) - \lambda(\mathbf{r})\sum_i n_i q_i^2 \phi(\mathbf{r})/kT = -\rho(\mathbf{r}) \qquad (2)$$

For biomolecules of arbitrary shape, the solution of equation (1) or (2) can only be obtained numerically, typically through finite-difference procedures. In this scheme, the PBE is discretized as follows

$$\begin{aligned}
&-h^{-2}\varepsilon_i(i-1,j,k)[\phi(i-1,j,k)-\phi(i,j,k)] \\
&-h^{-2}\varepsilon_i(i,j,k)\quad[\phi(i+1,j,k)-\phi(i,j,k)] \\
&-h^{-2}\varepsilon_j(i,j-1,k)[\phi(i,j-1,k)-\phi(i,j,k)] \\
&-h^{-2}\varepsilon_j(i,j,k)\quad[\phi(i,j+1,k)-\phi(i,j,k)] \\
&-h^{-2}\varepsilon_k(i,j,k-1)[\phi(i,j,k-1)-\phi(i,j,k)] \\
&-h^{-2}\varepsilon_k(i,j,k)\quad[\phi(i,j,k+1)-\phi(i,j,k)] \\
&+\quad \kappa^2 \quad\quad \phi(i,j,k) \quad\quad = h^{-3}q(i,j,k)
\end{aligned} \qquad (3)$$

where $h$ is the grid spacing in each dimension, $i$, $j$, and $k$ are the grid indexes along $x$, $y$ and $z$ axes, respectively. $\varepsilon_i(i,j,k)$ is the dielectric constant between grid points $(i,j,k)$ and $(i+1,j,k)$. $\varepsilon_j(i,j,k)$ and $\varepsilon_k(i,j,k)$ are defined similarly. All the related coefficients in Boltzmann term are absorbed into $\kappa^2$, and $q(i,j,k)$ is the charge within the cubic volume centered at $(i, j, k)$. The linear system can be conveniently written as

$$\mathbf{A}\phi = b, \qquad (4)$$

where $\mathbf{A}$ is the coefficient matrix of dielectric constants and the Boltzmann term, and $b$ is the constant vector of charges on the grids.

To solve equation (4), various solvers have been developed for biomolecular applications, such as successive over-relaxation (SOR),[15] conjugate gradient (CG),[15] (modified) incomplete

Cholesky conjugate gradient ((M)ICCG),[11] geometric multigrid (GMG),[16] and algebraic multigrid (AMG).[17] All solvers proceed from an initial guess of $\phi(i,j,k)$ to generate a sequence of improving solutions iteratively.

## 2.2 Conjugate Gradient Solvers

Symmetric and positive-definite linear systems are often solved with the CG solvers. The CG method searches for the exact solution along a series of conjugate directions, and is implemented as an iterative procedure as follows:

1. set $l=0$, $p_0 = r_0$
2. compute the norm of $\|r_l\|$. If $\|r_l\|/\|b\| < \delta$, output $\phi_l$. Otherwise go to the next step.
3. compute
$$\alpha_l = \frac{r_l^T p_l}{p_l^T \mathbf{A} p_l}, \quad \phi_{l+1} = \phi_l + \alpha_l p_l$$
4. compute
$$r_{l+1} = b - \mathbf{A}\phi_{l+1}, \quad \beta_l = \frac{r_{l+1}^T \mathbf{A} p_l}{p_l^T \mathbf{A} p_l}, \quad p_{l+1} = r_{l+1} + \beta_l p_l$$
5. set $l = l+1$ and go to step 2

The convergence of CG is optimal when the eigenvalues of the coefficient matrix are similar to each other.[11a] Thus preconditioner is often used in the CG method to achieve this goal. Specifically a preconditioner matrix $\mathbf{M}$ is introduced into equation (4)

$$(\mathbf{M}^{-1}\mathbf{A}\mathbf{M}^{-1})(\mathbf{M}f) = \mathbf{M}^{-1}b, \tag{5}$$

so that the new linear system becomes

$$\tilde{\mathbf{A}}\tilde{\phi} = \tilde{b} \tag{6}$$

By directly incorporating preconditioning into CG iteration, the resulting algorithm can be summarized as follows:

1. set $l=0$, $r_0 = b - A\phi_0$
2. solve $Mz_0 = r_0$ for $z_0$, let $p_0 = z_0$
3. calculate the norm of residue $\|r_l\|$. If $\|r_l\|/\|b\| < \delta$, output $\phi_l$. Otherwise go to the next step.
4. set $l = l+1$
5. calculate
$$\alpha_l = (r_{l-1}^T z_{l-1})/(p_{l-1}^T A p_{l-1}), \quad \phi_l = \phi_{l-1} + \alpha_l p_{l-1}, \quad r_l = r_{l-1} - \alpha_l A p_{l-1}$$
6. solve $Mz_l = r_l$ for $z_l$
7. calculate
$$\beta_l = (r_l^T z_l)/(r_{l-1}^T z_{l-1}), \quad p_l = z_l + \beta_l p_{l-1}$$
8. go to step 3

We can see that the preconditioned CG algorithm involves an additional operation at each iteration to solve the linear system $Mz_l = r_l$.

## 2.3 Incomplete Cholesky Preconditioners

A commonly used type of preconditioners is based on the incomplete $\mathbf{LDL}^T$ factorization

$$\mathbf{M} = (\tilde{\mathbf{D}} + \mathbf{L})\tilde{\mathbf{D}}^{-1}(\tilde{\mathbf{D}} + \mathbf{L}^T). \tag{7}$$

Here the matrices are related to the original coefficient matrix $\mathbf{A}$ as $\mathbf{A} = \mathbf{L} + \mathbf{D} + \mathbf{L}^T$ with $\mathbf{L}$ as the strictly lower triangular matrix of $\mathbf{A}$ and $\mathbf{D}$ as the positive diagonal matrix of $\mathbf{A}$. Finally $\tilde{\mathbf{D}}$ is an undetermined positive diagonal matrix. If the diagonal of $\mathbf{M}$ is defined as $\mathbf{D}$, the preconditioned conjugate gradient is termed ICCG. In MICCG, the diagonal elements of $\tilde{\mathbf{D}}$ are optimized to further improve the convergence.[8] The MICCG method is our default CPU implementation for our PBSA program in the Amber and AmberTools releases.

## 2.4 Jacobi Preconditioner

The Jacobi preconditioner (aka diagonal preconditioner) simply extracts the main diagonal **D** of **A** as **M**. Jacobi preconditioning is very inexpensive to use and is reasonably efficient for diagonally dominant matrices, though its reduction in the iteration number is modest. However, for the GPU implementation, the Jacobi preconditioner is advantageous because it is completely lack of row dependency, leading to great parallel efficiency. Additionally, the Jacobi preconditioner needs very little storage as to be discussed below.

## 2.5 Smoothed-aggregation-based Algebraic Multigrid Preconditioner

Multigrid methods are highly efficient techniques to solve linear or nonlinear equations. Typically there are two classes of multigrid methods: geometric multigrid (GMG) and algebraic multigrid (AMG).[18] GMG methods require prior physical/ mathematical knowledge of the underlying discretization and grid hierarchy, whereas AMG methods only require the coefficient matrix. Classical AMG methods involve the construction of a hierarchy of grids using the original coefficient matrix. The hierarchical grids are obtained by partitioning the grid nodes into coarse and fine grid nodes. The coarse grid nodes form a coarse level, and an interpolation operator, via a weighted sum of the coarse grid nodes, is used to interpolate a coarse level solution to a fine level. The restriction operator, usually taken as the transpose of the interpolation operator, is used to restrict a fine level solution to a coarse level.[19] Aggregation AMG methods obtain the hierarchical grids by aggregating a few fine grid nodes to form a coarse grid node. The interpolation operator uses a piecewise constant interpolation to obtain a fine level solution from a coarse level solution. This leads to rather sparse interpolation. The restriction operator is similar to that of the classical AMG methods. The aggregation scheme reduces the memory requirement and improves the interpolation efficiency, but it does not provide grid independent convergence.[19b] Therefore smooth interpolation or smooth aggregation is often used to improve the convergence.[20]

Unlike classical AMG, smoothed-aggregation-based AMG (SA-AMG) is not robust for various applications.[19b] Thus SA-AMG is often used as a preconditioner for generalized minimal residual and conjugate gradient methods. In this study, we tested the use of SA-AMG method to build a preconditioner (**M**) to the conjugate gradient method as implemented in CUSP.

**2.6 GPU Implementation**

The latest generations of GPU cards and Nvidia CUDA provide mature computing platforms for scientific applications. CUDA gives developers direct access to parallel computational elements (GPUs) and enables code to run concurrently in CPUs. Several CUDA-compatible libraries were utilized to implement a GPU-ready Amber/PBSA program. The CUDA Basic Linear Algebra Subroutines (cuBLAS) library is a GPU-accelerated BLAS library that are "6× to 17× faster" than the latest MKL.[21] The Nvidia CUDA Sparse Matrix (cuSPARSE) library provides basic linear algebra procedures for sparse matrix operations that are "up to 8× faster" than the latest MKL.[22] The cuSPARSE library is designed to interface with C or C++ functions. It supports multiple sparse matrix storage formats, such as Coordinate (COO), Compressed Sparse Row (CSR), Compressed Sparse Column (CSC), ELLPACK (ELL), Hybrid ELL+COO (HYB), and Blocked CSR. Finally CUSP is an open source C++ library based on Thrust. It can also provide sparse matrix operations in the CUDA environment.[23] CUSP supports COO, CSR, Diagonal (DIA), ELL, and HYB matrix formats.

**Table 1**. Details of tested solver combinations.

| Solver Combination | Description |
|---|---|
| Jacobi-DIA-CUSP | Jacobi-preconditioned CG using CUSP library with DIA matrix format |
| Jacobi-ELL-CUSP | Jacobi-preconditioned CG using CUSP library with ELL matrix format |
| Jacobi-HYB-CUSP | Jacobi-preconditioned CG using CUSP library with HYB matrix format |
| Jacobi-CSR-CUSP | Jacobi-preconditioned CG using CUSP library with CSR matrix format |
| Jacobi-COO-CUSP | Jacobi-preconditioned CG using CUSP library with COO matrix format |
| AMG-CSR-CUSP | Smoothed-aggregation-based AMG using CUSP library with CSR matrix format |
| AMG-COO-CUSP | Smoothed-aggregation-based AMG using CUSP library with COO matrix format |
| AMG-HYB-CUSP | Smoothed-aggregation-based AMG using CUSP library with HYB matrix format |
| CG-ELL-CUSP | CG using CUSP library with ELL matrix format |
| CG-HYB-CUSP | CG using CUSP library with HYB matrix format |
| CG-CSR-CUSP | CG using CUSP library with CSR matrix format |
| CG-COO-CUSP | CG using CUSP library with COO matrix format |
| CG-DIA-CUSP | CG using CUSP library with DIA matrix format |
| CG-CSR-cuSPARSE | CG using cuSPARSE library with CSR matrix format |
| ICCG-CSR-cuSPARSE | ICCG using cuSPARSE library with CSR matrix format |

In this study we implemented four types of FDM solvers, i.e. CG, ICCG, Jacobi-CG and SA-AMG-CG using cuBLAS, cuSPARSE, and CUSP libraries. We also tested these implementations with five different matrix formats DIA, CSR, COO, ELL, and HYB to analyze the impact of matrix formats upon efficiency. A total of 15 combinations are possible as summarized in Table 1. Apparently not every combination is available, e.g. the cuSPARSE library only works with the CSR format; Jacobi-CG only works with the CUSP library, and SA-AMG-CG only works with the CSR, COO and HYB formats in the CUSP library.

**2.7 Computational Details**

All CUDA solvers were implemented in the single precision within the Amber/PBSA program of the Amber 16 package.[24] A total of 573 biomolecular structures including proteins, short peptides, and nucleic acids in the Amber benchmark suite were used in our test.[31] These biomolecules consist of atoms ranging from 247 to 8,254 and have quite different geometries, and they were assigned charges of Cornell *et al*[25] and the modified bond radii.

All testings were performed with the following conditions unless specified otherwise. The convergence criterions of $10^{-3}$ and $10^{-6}$ were used for performance comparisons for high- or low-precision applications, respectively. The default grid spacing of 0.5 Å was used. The ratio of the grid dimension over the solute dimension was set to 1.5. No electrostatic focusing was applied for easy timing analysis. The potential values on all grids were initialized to zero. The dielectric constants were set to 80 and 1 for solvent and solute, respectively. The weighted harmonic average of the solvent and solute dielectric constants was used as the boundary dielectric constants. Therefore, the symmetric and positive-definite coefficient matrices were obtained and suitable for all tested linear solvers. In addition, the FDM matrix was initialized into CSR format and transformed into other formats when needed. Finally both the free space boundary condition (FBC) and periodic boundary condition (PBC) were tested. In PBC applications, we filled the matrix elements on the additional 6 bands into the original 7 bands and stored their column index non-consecutively in the CSR index arrays, thus we managed to use the same space as the original 7 bands in free boundary condition. All other parameters were set as default in the PBSA program in the Amber 16 package.[24]

We performed all measurements on a hybrid node with two NVIDIA GeForce GTX 980 Ti GPU cards and one Intel Xeon E5-1620 v3 CPU and 16GB main memory. The CPU timing measurements include all execution time of the core routine, i.e. time elapsed on both GPU and CPU, as well as time for transferring data between GPU and CPU.

## 3. Results and Discussion

### 3.1 Accuracy of GPU implementations

It is important to guarantee that the GPU implementations achieve consistent numerical results with existing CPU implementations within specified convergence criterion. As shown in Figure 1 for calculations in the free boundary condition, the electrostatic solvation energies on GPU (Jacobi-CG) and on CPU (CG) correlate quite well with both $10^{-3}$ and $10^{-6}$ convergence criteria. The linear regression slopes are 0.999931 and 0.999996, respectively, and the correlation coefficients are 1.0 for both. The maximum relative energy errors are $3.0\times10^{-3}$ and $6.3\times10^{-6}$, which are in agreement with the convergence criteria chosen.

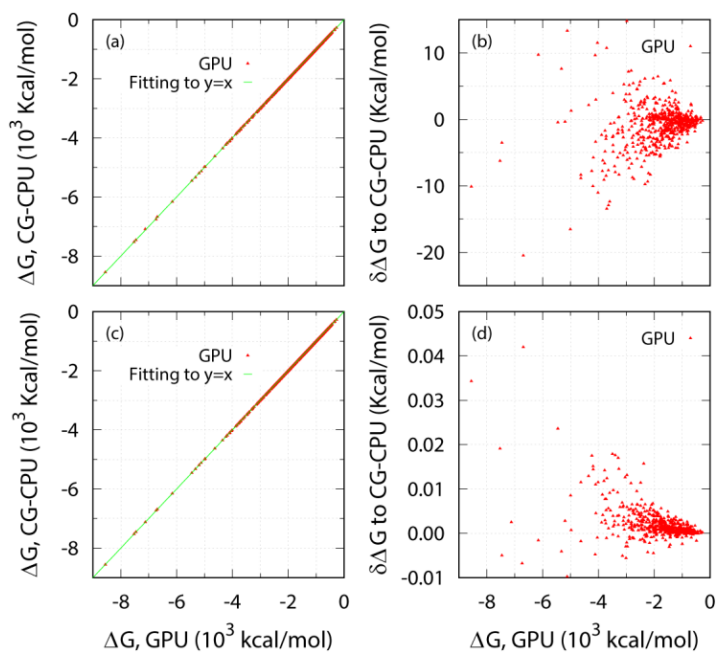

**Figure 1**. Correlations (a), (c) and differences (b), (d) of electrostatic solvation energies on GPU (Jacobi with CUSP library and DIA matrix format) and on CPU (CG) for the protein test set. Free space boundary condition was used. The convergence criterion was set to $10^{-3}$ (a), (b) and $10^{-6}$ (c), (d). The linear regression slopes are 0.999931 and 0.999996 for $10^{-3}$ and $10^{-6}$ criterion respectively, and the correlation coefficients are 1.0 for both.

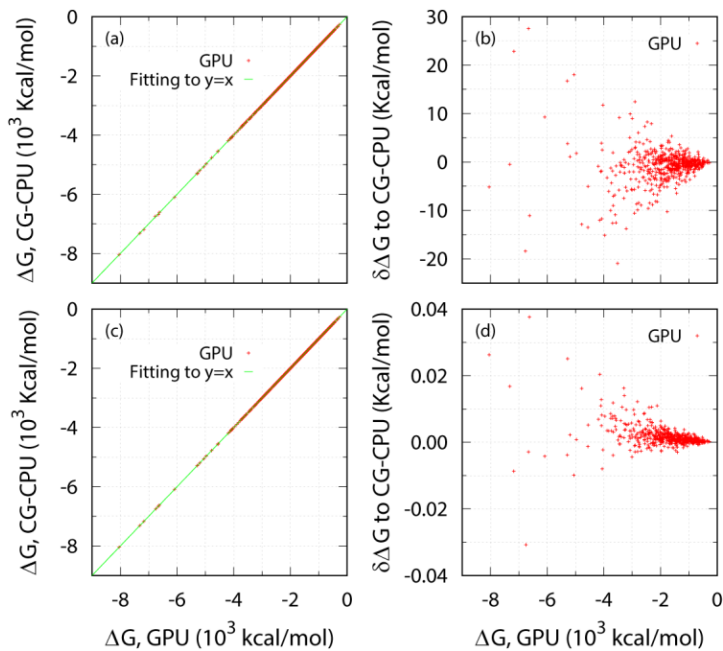

**Figure 2**. Correlations (a), (c) and differences (b), (d) of electrostatic solvation energies on GPU (Jacobi with CUSP library and DIA matrix format) and on CPU (CG) for the protein test set. The periodic boundary condition (PBC) was used. The convergence criterion was set to $10^{-3}$ (a), (b) and $10^{-6}$ (c), (d). The linear regression slopes are 0.999639 and 1.0 for $10^{-3}$ and $10^{-6}$ criterion respectively, and the correlation coefficients are 1.0 for both.

Similar agreements were also observed between the two implementations in the periodic boundary condition as shown in Figure 2. The linear regression slopes are 0.999639 and 1.0 for $10^{-3}$ and $10^{-6}$ convergence criteria, respectively, and the correlation coefficients are 1.0 for both. The maximum relative errors are $3.9 \times 10^{-3}$ and $5.8 \times 10^{-6}$, respectively, in agreement with the preset convergence criteria. For all other GPU implementations of which the data are not shown here, similar agreements were also observed.

### 3.2 Efficiency of GPU implementations

To compare the efficiency of GPU and CPU implementations, we first selected eight representative proteins and measured their solver CPU times with $10^{-3}$ and $10^{-6}$ convergence criteria, respectively. The standard CG solver as implemented on GPU and CPU was first analyzed. Table 2 shows that the GPU/CG solver overall performs better than the CPU/CG solver, with a speedup of ~6 to ~10 for the low convergence criterion and a speedup of ~8 to ~13 for the high convergence criterion. This is encouraging given that the tested GPU/CG solver is an unconditioned CG from the standard library without any change.

We next compared the GPU/CG solver with our default CPU solver CPU/ICCG, which was hand-optimized in the matrix-free fashion for modern CPUs. It is interesting to note that GPU/CG still performs better than CPU/ICCG for the larger proteins with the number of grid nodes over 2 million at the low convergence criterion. Furthermore, it performs better than CPU/ICCG for all tested proteins at the high convergence criterion.

A natural direction to go is to port ICCG to GPU, and this was implemented using the cuSPARSE library in this study. Unfortunately, our test shows that ICCG performs poorly on GPU platforms. This is consistent with the widely known fact that ICCG is not suitable for parallel platforms. Indeed, the inefficiency of the GPU/ICCG solver is significant: over 20 times slower than the standard CPU/CG solver. It should be pointed out that the the poor efficiency of GPU/ICCG is observed even with Nvidia's in-house optimization.[26] The specialized solver intends to find any independence in the sparse matrix during the analysis phase to solve the linear system in a parallel fashion.[26] In the case of linear PBE systems, however, this strategy fails to find any significant data independence in the seven-banded matrix.

**Table 2**. Average time (in second) used by CPU and GPU solvers for eight selected test proteins and representative solvers. The CG and Jacobi-preconditioned CG on GPU were carried out with CUSP library and DIA matrix format, and the SA-AMG-preconditioned CG on GPU was carried out with CUSP library and COO matrix format, while the ICCG solver on GPU was implemented with cuSPARSE library and CSR matrix format. The timing scheme for each solver include all execution time of the core routine code, i.e. time elapsed on device (GPU) and on host (CPU) and on transferring data between the device and the host. Both $10^{-3}$ and $10^{-6}$ criteria were used for comparison.

| | Protein | Ngrid | CPU | | GPU | | | |
|---|---|---|---|---|---|---|---|---|
| | | | CG | ICCG | CG | ICCG | Jacobi-CG | AMG-CG |
| Convergence $10^{-3}$ | 1pmc | 739431 | 1.81 | 0.21 | 0.32 | 24.65 | 0.26 | 0.57 |
| | 1e01 | 1010510 | 3.36 | 0.33 | 0.44 | 37.31 | 0.26 | 0.68 |
| | 1ghc | 1244220 | 5.02 | 0.46 | 0.54 | 51.33 | 0.31 | 0.79 |
| | 1f53 | 1466600 | 4.73 | 0.32 | 0.53 | 42.48 | 0.30 | 0.89 |
| | 1e0a | 1651190 | 4.47 | 0.47 | 0.52 | 60.57 | 0.32 | 0.97 |
| | 1ev0 | 1912380 | 5.48 | 0.79 | 0.61 | 91.38 | 0.41 | 1.06 |
| | 1dz7 | 2160050 | 7.62 | 0.86 | 0.72 | 106.19 | 0.43 | 1.15 |
| | 1ap0 | 2603130 | 6.75 | 1.22 | 0.67 | 144.31 | 0.51 | 1.34 |
| Convergence $10^{-6}$ | 1pmc | 739431 | 4.39 | 0.67 | 0.52 | 61.99 | 0.32 | 0.60 |
| | 1e01 | 1010510 | 7.28 | 0.96 | 0.69 | 84.98 | 0.36 | 0.72 |
| | 1ghc | 1244220 | 9.87 | 1.32 | 0.86 | 118.73 | 0.44 | 0.81 |
| | 1f53 | 1466600 | 12.68 | 1.47 | 1.05 | 129.50 | 0.47 | 0.92 |
| | 1e0a | 1651190 | 9.75 | 1.8 | 0.85 | 162.12 | 0.50 | 1.01 |
| | 1ev0 | 1912380 | 13.18 | 2.42 | 1.05 | 219.29 | 0.61 | 1.11 |
| | 1dz7 | 2160050 | 15.72 | 2.79 | 1.20 | 249.95 | 0.65 | 1.21 |
| | 1ap0 | 2603130 | 18.78 | 3.59 | 1.43 | 325.58 | 0.80 | 1.39 |

Nevertheless, there are other solvers that are more suitable for the parallel GPU platforms. It appears that several are available. Our comprehension analysis shows that Jacobi-CG is quite attractive. It was implemented with the CUSP library in the DIA matrix format. As shown in Table 2, the GPU/Jacobi-CG solver is about 7 to 13 times faster than CPU/CG for the low convergence criterion; and 11 to 23 times faster than CPU/CG for the high convergence criterion. The dramatically better performance of GPU/Jacobi-CG over GPU/ICCG lies in the simple utilization of the diagonal matrix as a preconditioner, which is completely without row dependency, so that it greatly facilitates parallel execution.

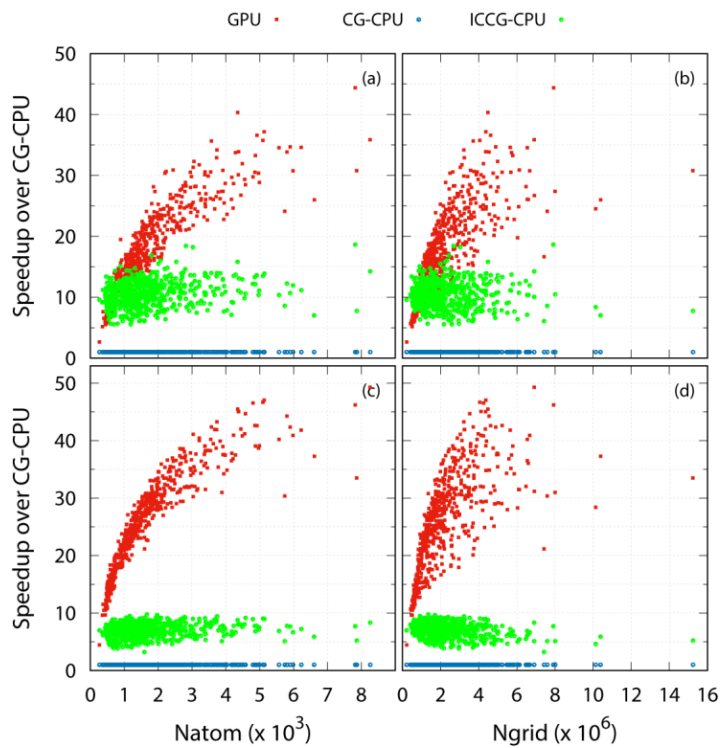

**Figure 3**. Comparison between PB solvers on GPU (Jacobi with CUSP library and DIA matrix format) and on CPU with free space boundary condition for the protein test set, as functions of number of atoms and grids respectively. The convergence criterion was set to $10^{-3}$ (a), (b) and $10^{-6}$ (c), (d).

Another interesting solver that can take advantage of GPU platforms is the SA-AMG-CG solver. We implemented the SA-AMG-CG solver in the CUSP library and observed reasonable speedup. Different from ICCG, the GPU/SA-AMG-CG implementation is observed to perform similarly among the best GPU implementations: slightly slower than GPU/CG and GPU/Jacobi-CG, but more efficient than both CPU/CG and CPU/ICCG at the high convergence criterion as shown in Table 2. It is less efficient than CPU/ICCG at the low convergence criterion, but clearly better than the standard CPU/CG implementation. These data indicate the potential to further implement multigrid types of linear solvers, such as geometric multi-grid solvers, for GPU platforms.

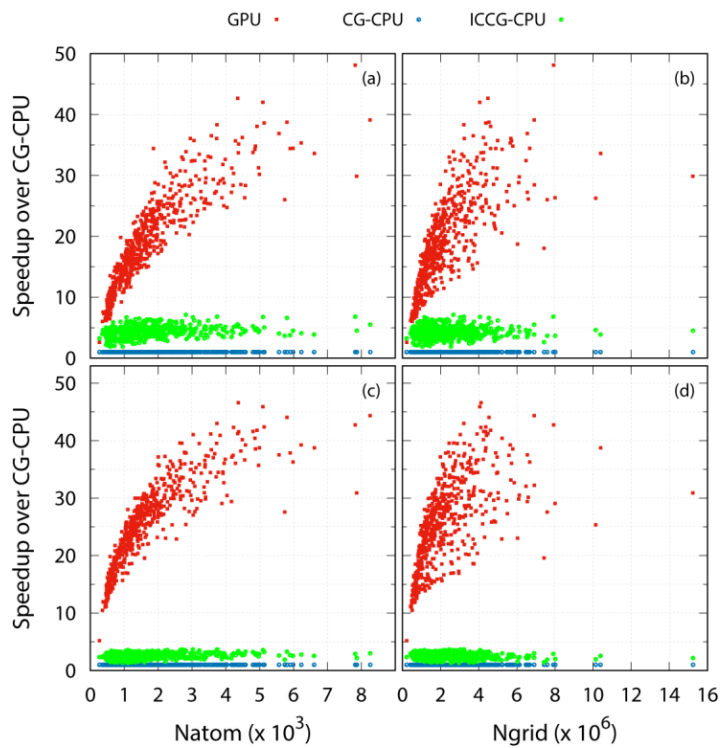

**Figure 4**. Comparison between PB solvers on GPU (Jacobi with CUSP library and DIA matrix format) and on CPU (ICCG) with PBC for the protein test set, as functions of number of atoms and grids respectively. The convergence criterion was set to $10^{-3}$ (a), (b) and $10^{-6}$ (c), (d).

Given the above detailed comparison of multiple implementations on selected proteins, it is clear that the GPU/Jacobi-CG is overall the most efficient implementation at both testing conditions (low and high convergence criteria). To properly gauge the overall speedups for typical applications for both free space boundary condition and periodic boundary condition, we plotted the speedup ratios of the GPU/Jacobi-CG implementation over the CPU/CG implementation using all test cases. As shown in Figure 3 for the free space boundary condition and Figure 4 for the periodic boundary condition, a speedup ratio of about 5 to 50 can be observed. The actual values clearly depend on the size/structure and of a given system. An interesting observation is that the speedup is not influenced much by the boundary conditions by comparing the trends in Figure 3 and Figure 4. However, it is clear that the CPU/ICCG implementation is more efficient in the free space boundary condition with speedup ratios up to 18 versus speedup ratios up to 6 in the periodic boundary condition in the low convergence criterion. This is because the difficulties in implementing periodic boundary condition in the highly optimized ICCG solver that prevent certainly data management ideas, i.e. array padding, to be used on CPU platforms as discussed previously.[3n]

As in other computational sciences, the sparse matrix structure is a typical feature when a partial differential equation, such as PBE, is discretized. As a result sparse matrix-vector multiplications (SpMV) are critical operations in PBE solvers and represents the dominant cost in many iterative methods. In our CPU/CG and CPU/ICCG implementations, hand tuning was employed to fully utilize the banded structure of the matrix for efficient SpMV operations. For example, the SpMV operation between boundary elements and the potential grids in the PBC linear solvers is carried out by directly shifting the column index into the array index, avoiding extra matrix column manipulation.[3n] In addition, with only several extra indices to mark the columns, seven arrays are enough to store the non-zero elements of the banded coefficient matrix, i.e. no extra row or column index is needed. Compared to the CSR format storage, this can save as much as 53% of the memory usage, which also leads to dramatically reduced memory load and store operations. However, these improvements in our CPU implementations are not fully available in the existing CUDA libraries. These features will be adopted when developing hand-optimized GPU solvers in our next step.

### 3.3 Other issues of GPU implementations

Efficiency of a GPU solver is also significantly affected by the matrix storage format. There are a number of sparse matrix representations with different storage requirements, computational characteristics, and methods of accessing and manipulating matrix elements as summarized in the Methods section. The DIA format is tailored for highly specific classes of matrices and is the most computationally attractive for the banded matrices in our linear PBE systems.[27] This is apparent in Figure 5 with Jacobi-DIA and CG-DIA being the best. However, the current cuSPARSE library does not support the DIA format, so that only the CSR format, a general-purpose format, was used for performance evaluation. Finally it should be pointed out none of the GPU/ICCG implementations are shown in Figure 5 due to their extremely long execution times.

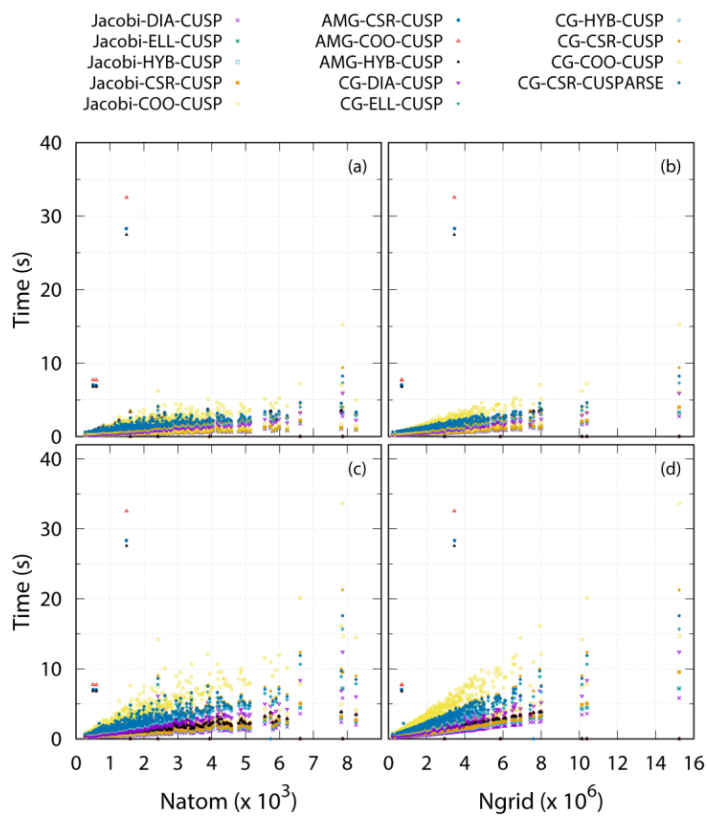

**Figure 5**. Average time used by different CPU and GPU solvers for selected test proteins. The detail of each solver combination is elaborated in Table 1. The GPU/ICCG solver is not listed due to their extremely long execution times. The convergence criterion was set to $10^{-3}$ (top) and $10^{-6}$ (bottom).

Given all the issues addressed, it is instructive to compare all GPU solver implementations as shown in Figure 5. This comparison also provides an opportunity to study the robustness of all GPU implementations. We examined all 573 test cases using both free space boundary condition and periodic boundary condition at both convergence criteria, to analyze the overall scaling of all GPU implementations. Indeed not every GPU implementation is robust enough to function properly for all tested conditions and molecules. There are five failures (failed to converge) and two unstable runs (converged but with unusually long time) for GPU/SA-AMG-CG in three out of the five tested formats (CSR/COO/HYB). The method fails in all test conditions and all molecules in the other two tested formats (DIA and ELL). Most failures were due to bad memory allocation, and others were due to unknown internal failures that lead to incorrect numerical solutions. This comprehensive scaling test confirms that GPU/Jacobi-CG from the CUSP library outperforms all its GPU and CPU counterparts significantly, and is also noticeably faster than our default CPU/ICCG implementation. As a reference, the CPU/ICCG is on average 10 times faster than CPU/CG as shown in Figure 3 and 4, consistent with the findings of Wang et al.[31]

**3.4 Memory usage of GPU implementations**

Memory usage is also crucial for GPU implementations because memory is often limited on most consumer-grade graphics cards. Apparently different solvers and matrix storage methods lead to different memory usages. In implementations with the CUSP library, typical memory usage is 88×Ngrid bytes for GPU/CG and 92×Ngrid for GPU/Jacobi-CG in the CSR matrix format. With the cuSPARSE library, GPU/CG consumes up to 76×Ngrid bytes of GPU memory and GPU/ICCG uses 180×Ngrid bytes in the CSR matrix. Here the estimations are based on the use of four-byte integer and float types. In addition, we managed to use about the same memory for both PBC and FBC applications as mentioned in Methods. This in part contributes to the consistent efficiency between the two boundary conditions for the tested molecules.

The above estimations are only based on those arrays explicitly allocated in the program. Run-time analysis by the NVIDIA hardware manage tool (nvidia-smi), however, shows that the total memory is about twice as much due the hidden buffer space allocated within the CUSP and cuSPARSE libraries. Thus the actual memory limit was underestimated in the estimations.

Extensive test of the fastest implementation, GPU/Jacobi-CG, shows that it was able to successfully complete linear PBE calculations with ~29.6 million grid nodes on the NVIDIA GTX 980 Ti cards with ~6GB GPU memory, about twice as smaller as the estimation.

## 4. Conclusions

In this study, we implemented multiple linear PBE solvers based on the standard CUDA libraries and conducted a systematic analysis on their performance with a large set of realistic biomolecules. We first analyzed the accuracy of the GPU implementation with respect to the CPU implementation in both free boundary condition and periodic boundary condition. The analysis shows that the GPU and CPU implementations agree within specified convergence criteria even if single precision was used in consumer grade graphics cards used in the test.

Many GPU solvers perform better than the standard CPU/CG solver, with various speedup ratios, depending on convergence criterion and size of the linear systems. In the comprehensive scaling test, our data shows that GPU/Jacobi-CG from the CUSP library outperforms all its GPU and CPU counterparts significantly. A speedup ratio of about 5 to 50 can be observed and it is not influenced much by the boundary conditions or convergence criteria. This should be compared with our default CPU implementation – the CPU/ICCG implementation, which is more efficient in the free space boundary condition. The speedup is reduced in the high convergence criterion in both boundary conditions tested. Unfortunately the ICCG method does not perform well on GPU platforms. Moreover, we implemented the SA-AMG-CG method and it was found to perform similarly among the best GPU implementations. These data indicate the potential to further implement multigrid types of linear solvers for GPU platforms.

It is also worth pointing out that the efficiency of a GPU solver is significantly affected by matrix storage formats. The DIA format is tailored for banded matrices and is the most computationally efficient for linear PBE matrices. Furthermore we discussed the memory usage of these solvers. Extensive test of the fastest implementation, GPU/Jacobi-CG, shows that it was able to successfully complete FDPB calculations with ~29.6 million grid points on the NVIDIA GTX 980 Ti cards with 6GB GPU memory, about twice as smaller as the theoretical analysis.

Finally further efficiency gain in GPU implementations is more likely to be achieved with customized matrix-free operations, integrated grid stencil setup on GPU, and also multigrid types of solvers, specifically tailored for our particular linear PBE problems. These developments are currently underway in our group.

## 5. Acknowledgements

This work was supported by National Institutes of Health/NIGMS (GM093040 & GM079383).